\documentclass[useAMS,usenatbib,dvipdfmx]{mn2e}
\usepackage{setspace, ulem, natbib, color, graphicx}
\title[Scratches of satellites]{Dynamical friction and scratches of orbiting satellite galaxies on host systems}
\author[Ogiya and Burkert]{Go Ogiya$^{1,2,3}$\thanks{E-mail:ogiya@mpe.mpg.de} and Andreas Burkert$^{1,2,3}$\thanks{Max--Planck Fellow}\\
$^{1}$Max-Planck-Institut f\"ur extraterrestrische Physik, Postfach 1312, Giessenbachstra\ss e, D-85741 Garching, Germany \\
$^{2}$Universit\"ats-Sternwarte M\"unchen, Scheinerstra\ss e 1, D-81679 M\"unchen, Germany\\
$^{3}$Excellence Cluster Universe, Boltzmannstr. 2, D-85748, Garching, Germany
}
\begin{document}

\date{Accepted 2016 January 8. Received 2016 January 8; in original form October 27}

\pagerange{\pageref{firstpage}--\pageref{lastpage}} \pubyear{2015}

\maketitle

\label{firstpage}

\begin{abstract}
We study the dynamical response of extended systems, hosts, to smaller systems, satellites, orbiting around the hosts using extremely high-resolution $N$-body simulations with up to one billion particles. 
This situation corresponds to minor mergers which are ubiquitous in the scenario of hierarchical structure formation in the universe. 
According to \cite{1943ApJ....97..255C}, satellites create density wakes along the orbit and the wakes cause a deceleration force on satellites, i.e. dynamical friction. 
This study proposes an analytical model to predict the dynamical response of hosts 
as reflected in their density distribution 
and finds not only traditional wakes but also mirror images of over- and underdensities centered on the host. 
Our controlled 
$N$-body simulations with high resolutions verify the predictions of the analytical model.
We apply our analytical model to the expected dynamical response of nearby interacting galaxy pairs, the Milky Way - Large Magellanic Cloud system and the M31 - M33 system. 
\end{abstract}


\section{Introduction}
\label{sec:int}
\cite{1943ApJ....97..255C} first discussed a fundamental physical process called dynamical friction for collisionless systems. 
According to his calculation, a massive object moving through a sea of particles produces density enhancements, `wakes' behind itself due to its gravitational force and the gravitationally induced wakes generate a decelerating force on the moving massive object. 
As a consequence, the massive object loses its orbital energy and angular momentum. 
\cite{1999ApJ...513..252O} later on studied the process of dynamical friction for gaseous systems. 

Dynamical friction arises in various astronomical phenomena \citep{2008gady.book.....B}. 
Considering galaxy mergers, satellite galaxies orbiting around larger ones lose orbital energy and angular momentum by the effects of dynamical friction and eventually fall into the center of their hosts 
\citep[e.g.][]{1976MNRAS.174..467W, 1999ApJ...525..720C, 2001ApJ...559..716T}.
A prominent example is the orbit decay of the Large Magellanic Cloud that has been studied in great details 
\citep[e.g][]{1976ApJ...203...72T, 1980PASJ...32..581M, 1994MNRAS.266..567G, 2003ApJ...582..196H}. 
Dynamical friction also plays important roles for the formation and evolution of black hole binaries 
\citep{1980Natur.287..307B, 2002MNRAS.331..935Y, 2004ApJ...602...93M, 2011ApJ...728L..31T, 2013ApJ...777L..14F}. 
Studies based on numerical simulations reveal the validity but also 
problems of Chandrasekhar's theory \citep{2006PASJ...58..743F,2006MNRAS.373.1451R,  2009MNRAS.397..709I}. 

Important insight has been gained, applying the arguments of dynamical friction to flattened systems, such as galactic stellar discs. 
\cite{1977MNRAS.181..735B} proposed a correction of Chandrasekhar's formulation of dynamical friction for systems with anisotropic velocity distributions. 
\cite{2004MNRAS.349..747P} confirmed the reliability of Binney's treatment numerically. 
Dynamical friction generates the heating source to explain the thickness of stellar discs \citep[e.g.][]{1999MNRAS.304..254V, 2008ApJ...674L..77M}. 
\cite{2002MNRAS.333..779P} and \cite{2008MNRAS.389.1041R} found that orbiting satellite galaxies around a larger galaxy with disc components are dragged into the disc plane by their dynamical friction. 
This may explain the distribution of satellite galaxies around the Milky Way \citep{2002MNRAS.333..779P} and predicts the existence of stable disc structures of dark matter which significantly boost capturing dark matter particles \citep{2008MNRAS.389.1041R, 2009MNRAS.397...44R}. 

Gravitational wakes 
were investigated in details 
analytically by early work \citep[e.g.][]{1983A&A...117....9M, 1989MNRAS.239..549W}.  
However, only a few studies have confirmed gravitational wakes directly by using numerical simulations in spite of the importance 
to get a better understanding 
of dynamical friction. 
\citet{2002ApJ...580..627W, 2007MNRAS.375..425W} 
showed 
over- and underdensities induced by bars around the center of galaxies. 
\cite{2012ApJ...745...83A} 
illustrated 
the density response induced by black holes in galactic nuclei. 
For gaseous systems, \cite{2009ApJ...703.1278K} and \cite{2010ApJ...725.1069K} 
analysed 
density wakes by hydrodynamic simulations. 

Previous numerical studies were limited by particle resolution. 
The motivation of this study is to investigate the features of scratches induced by orbiting smaller systems (satellites) on larger ones (hosts) in great details adopting one billion particles. 
The structure of this paper is as follows. 
Section \ref{sec:ana} provides an analytical description to predict the response of the hosts 
as reflected in their density distribution. 
We find not only traditional gravitational wakes, but also a mirror image of over- and underdensities around the center of hosts. 
In Section \ref{sec:sim}, we perform high-resolution $N$-body simulations to test the analytical prediction. 
The results of our simulations well match to the predictions of the analytical description.  
We use the analytical model for the galaxy pairs, the Milky Way - Large Magellanic Cloud system and the M31 - M33 system as applications in Section \ref{sec:app}. 
Section \ref{sec:sum} summarizes and discusses the results.

\section{Analytical model}
\label{sec:ana}
The formula of dynamical friction proposed by \cite{1943ApJ....97..255C} can be written as
\begin{equation}
\frac{d {\bf v}_{\rm sat}}{dt} = -4 \pi G^2 M_{\rm sat} \ln{\Lambda} \rho_{\rm host} \frac{v_{\rm sat}}{v_{\rm sat}^3}, \label{eq:df_ori}
\end{equation}
where $M_{\rm sat}$ and $\rho_{\rm host}$ 
are 
the mass of the satellite and the density of the host, and ${\bf v}_{\rm sat}$ represents the velocity of the satellite in the rest frame of the host. 
$G$ and $\ln{\Lambda}$ are the gravitational constant and the Coulomb logarithm, respectively.
Here, the host systems are assumed to have an isotropic velocity field in the initial equilibrium state. 
We suppose that the mass of a particle which belongs to the host is negligible compared with $M_{\rm sat}$ and assume that all particles in the host contribute to the force of dynamical friction for simplicity although only host particles which satisfy $v < |{\bf v}_{\rm sat}|$ cause dynamical friction in the original Chandrasekhar's theory.
These assumptions do not influence our conclusion as shown in Section \ref{sec:sim}. 
In Chandrasekhar's derivation of dynamical friction, an infinite homogeneous 
particle 
distribution is assumed. 
However, galaxies and dark haloes have finite sizes and gradients in their density structures. 
Thus the Coulomb logarithm may have a dependence on the satellite position within the host \citep[e.g.][]{2003ApJ...582..196H, 2005A&A...431..861J}.
One may modify Equation (\ref{eq:df_ori}) as follows, 
\begin{equation}
\frac{d {\bf v}_{\rm sat}}{dt} = -4 \pi G^2 M_{\rm sat} \ln{\Lambda({\bf r}_{\rm sat})} \rho_{\rm host}({\bf r}_{\rm sat}) \frac{v_{\rm sat}}{v_{\rm sat}^3}, \label{eq:df_mod}
\end{equation}
 where ${\bf r}_{\rm sat}$ is the position of the satellite in the frame whose origin is the center of the host. 
Hereafter, we study the evolution in this rest frame.

Orbiting satellites are perturbers for hosts. 
Because of the gravitational force of the satellite, induced density perturbations will arise in the host. 
We label physical quantities of the host in the equilibrium state as `0' and induced ones as `1', respectively, i.e. 
\begin{eqnarray} 
\rho_{\rm host} = \rho_0 + \rho_1, \\
\Phi_{\rm host} = \Phi_0 + \Phi_1,  
\end{eqnarray}
where $\Phi$ 
is 
the gravitational potential. 
Poisson's equation connects the gravitational potential with the density field. 
For the induced quantities, we get 
\begin{equation}
{\bf \nabla}^2 \Phi_1 = -{\bf \nabla} \cdot {\bf g}_1 = 4 \pi G \rho_1, \label{eq:poisson}
\end{equation}
where ${\bf g}_1$ is the specific gravitational force caused by the induced density perturbations. 
We can regard the deceleration force of the dynamical friction process as ${\bf g}_1$. 
Substituting Equation (\ref{eq:df_mod}) into Equation (\ref{eq:poisson}), the induced density field is 
\begin{eqnarray}
\rho_1({\bf r}, {\bf v}_{\rm sat}) =  G M_{\rm sat} [ \rho_0({\bf r})  {\bf \nabla} \ln{\Lambda({\bf r})} + \ln{\Lambda({\bf r})} {\bf \nabla} \rho_0({\bf r})] \cdot \frac{{\bf v}_{\rm sat}}{v_{\rm sat}^3}. \label{eq:ind_dns}
\end{eqnarray}
Here, we assume that the absolute values of induced quantities are much smaller than those in the equilibrium state. 

We need to provide the density distribution of the host in the equilibrium state, $\rho_0({\bf r})$ and the Coulomb logarithm, $\ln{\Lambda({\bf r})}$ to use Equation (\ref{eq:ind_dns}). 
The density distribution of the host galaxy may be expressed well by the following double power-law formula, 
\begin{equation}
\rho_0(r) = \frac{\rho_{\rm s}}{r^{\alpha} [1 + (r / r_{\rm s})]^{\beta}}, \label{eq:double_power_law}
\end{equation}
where $\rho_{\rm s}$ and $r_{\rm s}$ are the scale density and length, respectively. 
The model of $\alpha=1, \beta=2$ is known as the Navarro-Frenk-White profile which well matches the density structure of cold dark matter haloes obtained in dissipationless cosmological simulations \citep{1997ApJ...490..493N}. 
The Hernquist profile \citep{1990ApJ...356..359H} which reproduces the de Vaucouleurs law \citep{1948AnAp...11..247D} and is frequently used to expess density structures of elliptical galaxies or bulges, corresponds to the model of $\alpha=1, \beta=3$ \citep[see however e.g.][]{1995ApJ...447L..25B}. 
The derivative of Equation (\ref{eq:double_power_law}) is given by 
\begin{equation}
{\bf \nabla} \rho_0 (r) = -\rho_0(r) \biggl [ \frac{\alpha}{r} + \frac{\beta}{r + r_{\rm s}} \biggr] \frac{{\bf r}}{r}. \label{eq:dns_grad}
\end{equation}

The Coulomb logarithm is defined as $\ln{\Lambda} \equiv \ln{(b_{\rm max} / b_{\rm min})}$ where $b_{\rm max}$ and $b_{\rm min}$ are the maximum and minimum impact parameters, respectively. 
The maximum impact parameter, $b_{\rm max}$ should depend on the vector pointing from the satelllite to given points, $ {\bf r}$, ${\bf d} = {\bf r} - {\bf r}_{\rm sat}$ since $b_{\rm max}$ determines the region affected by the gravitational force of the satellite. 
For simplicity, we suppose that $b_{\rm min}$ is a constant, 
\begin{equation}
b_{\rm min} = A l, \label{eq:bmin}
\end{equation}
where $A$ and $l$ mean a constant and the size of the satellite, respectively \citep[see also][]{1976ApJ...203...72T, 2003ApJ...582..196H}. 

We define the maximum impact parameter, $b_{\rm max}$ by the following procedure. 
Let us consider a plane which is perpendicular to the velocity vector of the satellite, ${\bf v}_{\rm sat}$. 
The vector, ${\bf v}_{\rm sat} \Delta t$ measures the distance between the satellite and plane and determines the plane on which a point is given by ${\bf r}_{\rm sat} - {\bf v}_{\rm sat} \Delta t$. 
Given points, ${\bf r}$ on the plane satisfy the condition, 
\begin{equation}
({\bf r} - {\bf r}_{\rm sat} + {\bf v}_{\rm sat} \Delta t) \cdot {\bf v}_{\rm sat} = 0. \label{eq:ip}
\end{equation}
From Equation (\ref{eq:ip}), $\Delta t$ is derived by 
\begin{equation}
\Delta t = -\frac{{\bf v}_{\rm sat} \cdot ({\bf r} - {\bf r}_{\rm sat})}{v_{\rm sat}^2} = -\frac{d \cos{\phi}}{v_{\rm sat}} \label{eq:dt}
\end{equation}
where $\phi$ is the angle between the vectors, ${\bf d}$ and ${\bf v}_{\rm sat}$. 
We define $b_{\rm max}$ as the length of a free-fall motion in the time interval, $\Delta t$, i.e. $b_{\rm max} = (1/2) a \Delta t^2$, behind the satellite at which ${\bf v}_{\rm sat} \cdot {\bf d} < 0$ is satisfied. 
The absolute value of the gravitational acceleration in the perpendicular direction to ${\bf v}_{\rm sat}$, $a$ is 
\begin{equation}
a = \frac{G M_{\rm sat}}{[b_{\rm max}^2 + (v_{\rm sat} \Delta t)^2]^{3/2}} b_{\rm max}. \label{eq:a_para}
\end{equation}
For a simple evaluation of the gravitational acceleration we assume that the satellite is located at 
${\bf r}_{\rm sat}$ from $t=T - \Delta t$ to $t = T$. 
Solving Equation (\ref{eq:a_para}) for $b_{\rm max}$, we obtain 
\begin{equation}
b_{\rm max} = \sqrt{\biggl [ \frac{G^2 M_{\rm sat}^2 \Delta t^4}{4} \biggr ]^{1/3} - (v_{\rm sat} \Delta t)^2}. \label{eq:bmax}
\end{equation} 
for points at which ${\bf v}_{\rm sat} \cdot {\bf d} < 0$ is satisfied. 
Also, 
\begin{equation}
b_{\rm max} =  B b_{\rm min} \label{eq:bmax2}
\end{equation}
for points at which ${\bf v}_{\rm sat} \cdot {\bf d} \geq 0$ is satisfied. 
Here, $B$ is a constant. 
The derivative is  
\begin{equation}
{\bf \nabla} \ln{\Lambda({\bf r})} = \frac{1}{b_{\rm max}^2} \biggl [ \biggl ( \frac{2 G^2 M_{\rm sat}^2 \cos^4{\phi}}{27 v_{\rm sat}^4} d \biggr )^{1/3} - d \cos^2{\phi} \biggr ] \frac{{\bf d}}{d} \label{eq:bmax_grad}
\end{equation}
for points at which ${\bf v}_{\rm sat} \cdot {\bf d} < 0$ and 
\begin{equation}
{\bf \nabla} \ln{\Lambda({\bf r})} = 0
\end{equation}
for points at which ${\bf v}_{\rm sat} \cdot {\bf d} \geq 0$ is satisfied, respectively. 
When $b_{\rm max} < B b_{\rm min}$ in Equation (\ref{eq:bmax}), we set $b_{\rm max} = B b_{\rm min}$ and ${\bf \nabla} \ln{\Lambda({\bf r})} = 0$. 
Hence, a constant, $B$ defines the minimum value of the Coulomb logarithm.  

Combined with Equations (\ref{eq:ind_dns}) and (\ref{eq:bmax_grad}), the first term in Equation (\ref{eq:ind_dns}) generates density perturbations around the satellite. 
Since Equation (\ref{eq:bmax}) defines $b_{\rm max}$ on the back side of the satellite motion, the induced density arises only behind the satellite. 
Hence, the first term represents Chandrasekhar's original gravitational wake. 
As indicated by combining Equations (\ref{eq:ind_dns}) and (\ref{eq:dns_grad}), the sign of the density fluctuations caused by the second term of Equation (\ref{eq:ind_dns}) depends on the angle between the vectors of the satellite velocity and the position. 
For spherical systems, the density fluctuations are symmetric with respect to the center of the host. 

\section{Simulations}
\label{sec:sim}

\subsection{Set up}
\label{sec:setup}
\begin{table}
\begin{center}
\caption{Summary of simulation runs. 
Description of columns: 
(1) Name of simulation runs. 
(2) Mass ratio between the host and satellite, $M \equiv M_{\rm v, host} / M_{\rm v, sat}$. 
(3) Total number of particles, $N_{\rm tot} = N_{\rm host} + N_{\rm sat}$. 
}
\begin{tabular}{ccccccc}
Run & $M$ & $N_{\rm tot}$ \\ 
(1) & (2) & (3) \\
\hline 
A & 100.0 & 105,906,176 \\
B & 50.0  &  106,954,752 \\
C & 20.0  &  110,100,480 \\
D & 10.0  &  115,343,360 \\
E & 10.0  &  1,038,090,240 \\
\hline 
\end{tabular}
\label{tab:run}
\end{center}
\end{table}

We perform controlled collisionless $N$-body simulations with a parallelized code optimized for graphic processing unit (GPU) clusters. 
The numerical code employs the tree algorithm proposed by \cite{1986Natur.324..446B} and the second-order Runge-Kutta scheme in time integration. 
Along the lines of \cite{2011arXiv1112.4539N}, CPU cores construct tree structures of particles and GPU cards compute the gravitational force among particles through traversing the tree structures \citep{2013JPhCS.454a2014O}. 

We simulate mergers between two systems, the host and 
the 
satellite. 
In order to generate $N$-body systems which follow the NFW density profile, i.e. $\alpha=1, \beta=2$ in Equation (\ref{eq:double_power_law}), in the equilibrium states, we use the method proposed by \cite{2006ApJ...641..647K} \citep[see also][]{1916MNRAS..76..572E}. 
The phase-space distribution function is assumed to depend only on energy and the systems have 
an 
isotropic velocity dispersion initially. 
The distribution of particles is truncated at the virial radius, $R_{\rm v}$ 
that is related to the virial mass, 
$M_{\rm v}$ 
\begin{equation}
M_{\rm v} = \frac{4 \pi}{3}  \Delta \rho_{\rm crit} (1+z)^3 R_{\rm v}^3, \label{eq:rv}
\end{equation}
where $\rho_{\rm crit}$ and $z$ are the critical density of the universe and redshift, respectively. 
We adopt a conventional value of the overdensity, $\Delta = 200$ in this study. 
The concentration parameter, $c$ is defined by $c = R_{\rm v} / r_{\rm s}$. 
The host and satellite have $c = 10$ and $15$, respectively. 

The initial separation between the centers of the host and satellite is the virial radius of the host, $R_{\rm v,host}$. 
The orbit has circularity $\eta = 0.5$ initially. 
This value is very typical, as suggested by cosmological $N$-body simulations \citep{2006A&A...445..403K}. 
The initial velocity of the satellite is $V_{\rm ini} = \eta V_{\rm c}(R_{\rm v,host}) = \eta (GM_{\rm v, host} / R_{\rm v,host})^{1/2}$. 
Here, $V_{\rm c}(r)$ 
is the 
circular velocity measured at $r$.  
In the coordinate system, centered on the host, the initial position and velocity vectors of the satellite are ${\bf X} = (R_{\rm v,host}, 0, 0)$ and ${\bf V} = (0, V_{\rm ini}, 0)$, respectively. 

The position and velocity of the host and satellite are determined by the following procedure. 
We calculate the bound mass of each system with the method proposed by \cite{1993PASJ...45..289F}. 
Particles initially belong either to the host or to the satellite system. 
At each snapshot, we compute the gravitational potential of each particle by particles which are still bound to the system at the previous snapshot. 
The bulk velocity of each system is determined by an iterative procedure \citep{2006PASJ...58..743F}. 
When the binding energy of a particle is positive, the particle is regarded as an escaper. 
We define the center of mass and bulk velocity of bound particles as the position and velocity of each system at given time. 

We construct the host systems which have the virial mass, $M_{\rm v, host}$ with $N_{\rm host}$ particles. 
For satellite systems with a virial mass, $M_{\rm v, sat}$, $N_{\rm sat} = (M_{\rm v, sat} / M_{\rm v, host}) N_{\rm host}$ particles are employed. 
Hence, all particles have equal masses in each simulation and the total number of particles is $N_{\rm tot} = (1 + M_{\rm v, sat} / M_{\rm v, host}) N_{\rm host}$. 
Table \ref{tab:run} summarizes the simulations. 
Because we use sufficient numbers of particles, artificial effects such as two-body relaxation are negligible. 
In the simulations, 
the softening length is $\epsilon = 4 R_{\rm v, host} / \sqrt{N_{\rm host}}$ and the tolerance parameter of the tree algorithm is $\theta = 0.6$ \citep{2003MNRAS.338...14P}. 

We are free to scale the mass, length and timescales since our simulations only take into account gravitational effects. 
For Milky Way sized haloes with 
$M_{\rm v, host} = 10^{12} M_{\rm \odot}$ and cosmic redshifts, $z = 0$, the virial radius of the host halo is $R_{\rm v, host} \approx 211 {\rm kpc}$ and the dynamical time of the host, $T_{\rm d, host}$ defined by 
\begin{equation}
T_{\rm d,host} \equiv \frac{R_{\rm v,host}}{V_{\rm c}(R_{\rm v,host})} = \sqrt{\frac{R_{\rm v,host}^3}{G M_{\rm v,host}}}, \label{eq:td}
\end{equation}
is $\approx 1.45 {\rm Gyr}$. 
Here, a Hubble constant of $H_0 = 67.5 {\rm km/s/Mpc}$ \citep{2015arXiv150201589P} is adopted. 
Simulation data are output every $0.05T_{\rm d, host}$. 

\subsection{Results}
\label{sec:res}
\begin{figure}
  \centering 
   \includegraphics[width=75mm]{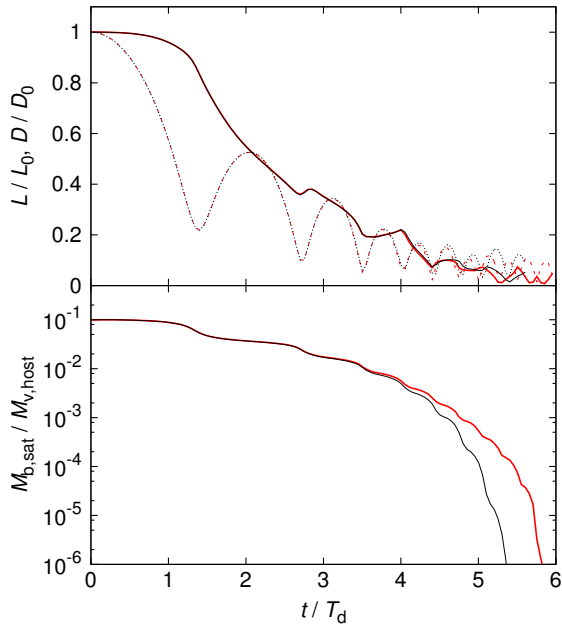}
     \caption{
       Time evolution of the satellite orbit (upper panel) and bound mass of the satellite, $M_{\rm b,sat}$ (lower panel) derived from 
run D (black) and E (red). 
       In the upper panel, solid and dashed lines represent the specific angular momentum, $L$ of the satellite orbit relative to the host and the distance, $D$ between the centers of the host and satellite. 
       $L$ and $D$ are normalized by the initial values, $L_0$ and $D_0$. 
       $M_{\rm b,sat}$ and time, $t$ are scaled by the virial mass of the host, $M_{\rm v,host}$ and dynamical time, $T_{\rm d, host}$, respectively. 
       \label{fig:evo}
     }
\end{figure}

\begin{figure*}
  \centering 
  \includegraphics[width=120mm]{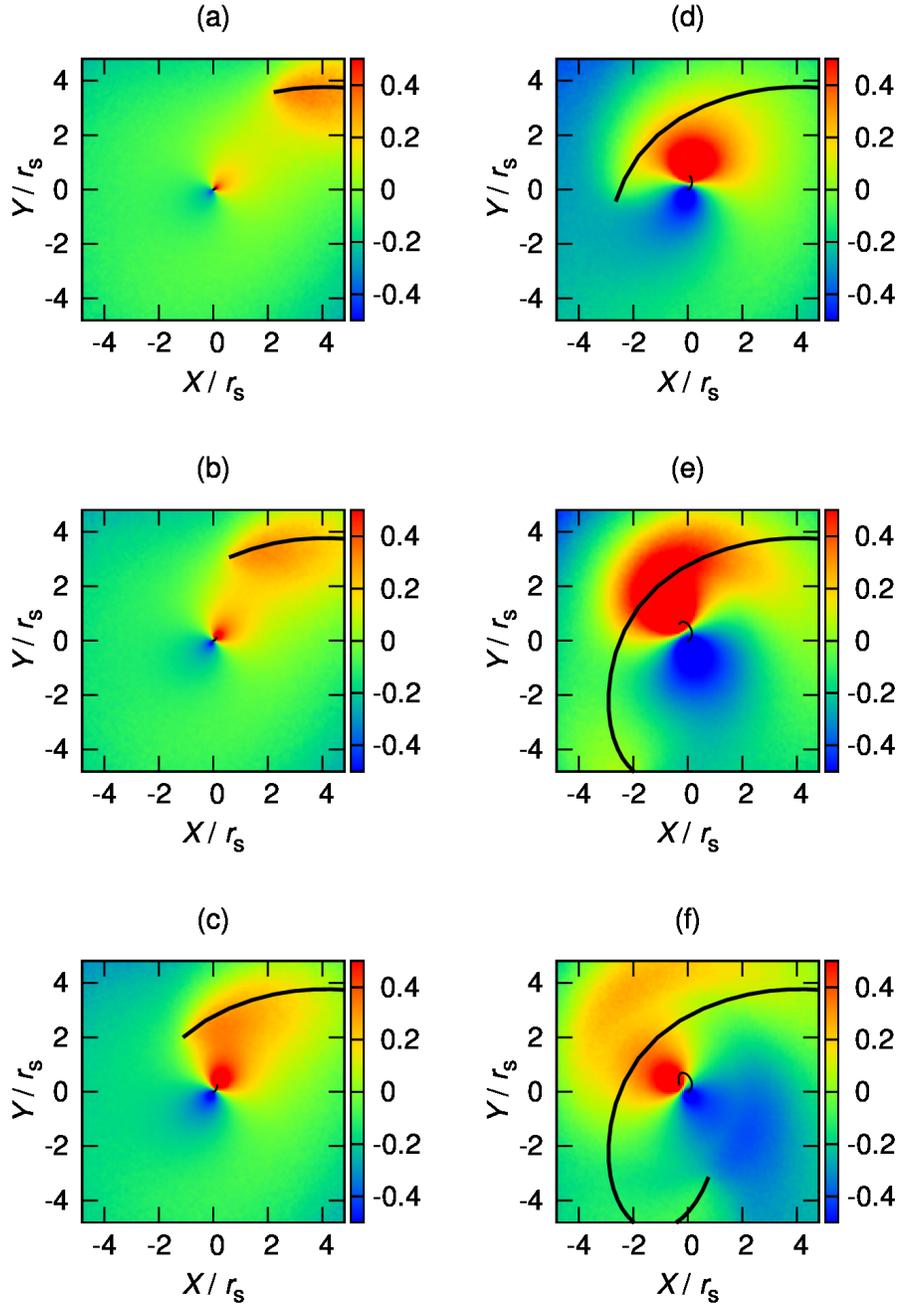}
     \caption{
       Distribution of enhancement and reduction in the column density of the host system derived from run E. 
       The color bar represents enhancement and reduction in the column density, $(\Sigma - \Sigma_0) / \Sigma_0$, where $\Sigma_0$ is the initial column density at given position in the host frame. 
       Spatial coordinates are scaled by the scale length of the host system, $r_{\rm s}$. 
         Thick and thin black lines show the satellite orbit and motion of the minimum potential point in the center-of-mass frame. 
       Panels (a), (b), (c), (d), (e) and (f) demonstrate snapshots at $T = 1.15, 1.25, 1.35, 1.5, 2.0$ and $2.5 T_{\rm d,host}$, respectively. 
       \label{fig:map1}
     }
\end{figure*}

\begin{figure*}
  \centering 
  \includegraphics[width=120mm]{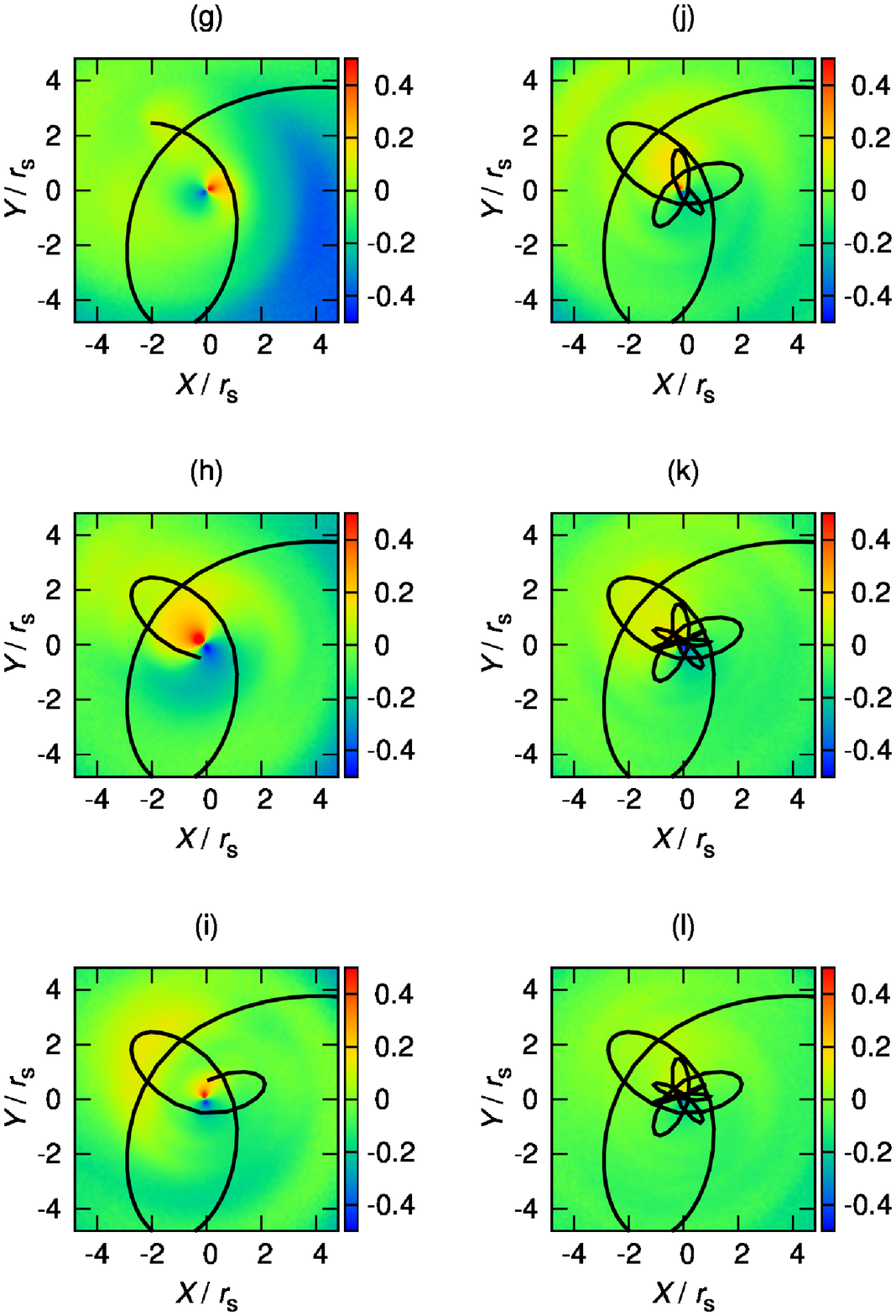}
     \caption{
       Same as Figure \ref{fig:map1}. 
       Panels (g), (h), (i), (j), (k) and (l) demonstrate snapshots at $T = 3.0, 3.5, 4.0, 5.0, 6.0$ and $7.0 T_{\rm d,host}$, respectively. 
         The thin black line which 
       represents the motion of the minimum potential point in the center-of-mass frame is not drawn for 
         better 
       visibility.
       \label{fig:map2}
     }
\end{figure*}

Figure \ref{fig:evo} summarizes a typical case of galaxy merger. 
The satellite loses angular momentum due to dynamical friction and the orbit shrinks gradually (upper panel). 
In the meantime, the satellite is stripped by the tidal force of the host, especially when it approaches the pericenters. 
Eventually, it is completely destroyed at 
$t = 5.65 T_{\rm d,host}$ in run D and $6.0 T_{\rm d,host}$ in run E, respectively (lower panel). 
Figure \ref{fig:evo} also tests the numerical convergence of 
the 
simulations. 
The orbital evolution is well converged. 
The evolution of the bound mass of the satellite 
for both resolutions deviates only at $t > 4T_{\rm d,host}$, at which most of its mass has already been stripped away and significant density scratches do not arise (see Figures \ref{fig:map1} and \ref{fig:map2}). 
Hence, the results of this paper do not depend on the number of particles, $N$. 
As shown in this figure, dynamical friction plays a key role during the merging process. 

Figures \ref{fig:map1} and \ref{fig:map2} illustrate that the orbiting satellite leaves significant scratches on the host. 
We find two kinds of scratches in 
its 
density distribution. 
The first one is the gravitational wake which has been found and discussed by many previous studies. 
The gravitational wake arises along the satellite orbit as imagined by \cite{1943ApJ....97..255C}. 
It can be found more clearly in the early phase of the merger process [see panels (a) and (b)] since it mixes with another type of scratch described below in the later phase. 

The second type of scratch caused by the gravitational field of the satellite is a pair of density 
enhancements and reductions 
around the center of the host. 
Similar results have been obtained analytically by \cite{1989MNRAS.239..549W}. 
This distribution of over- and underdensities is mirror image-like and the mirror plane is roughly perpendicular to the direction of the velocity vector of the satellite. 
The underdensity is located in the direction of the velocity vector of the satellite and the overdensity arises in the opposite direction. 
The directions are more visible in the early phase again due to the mixing of both, the wake and the central perturbation in the later phase. 

This central dipole scratch is caused by the motion of the location of the minimum of the potential, i.e. highest density point in the initial state. 
The tidal force of the satellite halo perturbs the position of the highest density point and it well matches the position of the 
highest 
overdensity. 
Hence, the motion of the minimum potential point well traces the overdensity around the center of the host in the center-of-mass frame (see thin black line in Figure \ref{fig:map1}). 
Assuming that the satellite is a less massive particle in a two-body problem, the thin line looks like an orbit of 
the corresponding more massive particle. 
Because of mass conservation, the overdensity on one side causes a corresponding underdensity on the opposite side of the minimum potential point with respect to the center of mass in the host. 
Since there is a single point 
where the potential has its minimum 
in the host, the central over- and underdensities have a dipole structure. 
The effect does not affect the bulk structure of the host and it retains the initial spherical shape on the whole. 

Figures \ref{fig:map1} and \ref{fig:map2} also show that the amplitude of over- and underdensities decreases with time. 
This is because of the decreasing satellite mass as a result of tidal stripping [see Equation (\ref{eq:ind_dns})]. 
After tidal disruption of the satellite, little scratches remain for some time [panels (k) and (l)]. 

\begin{figure}
  \centering 
   \includegraphics[width=85mm]{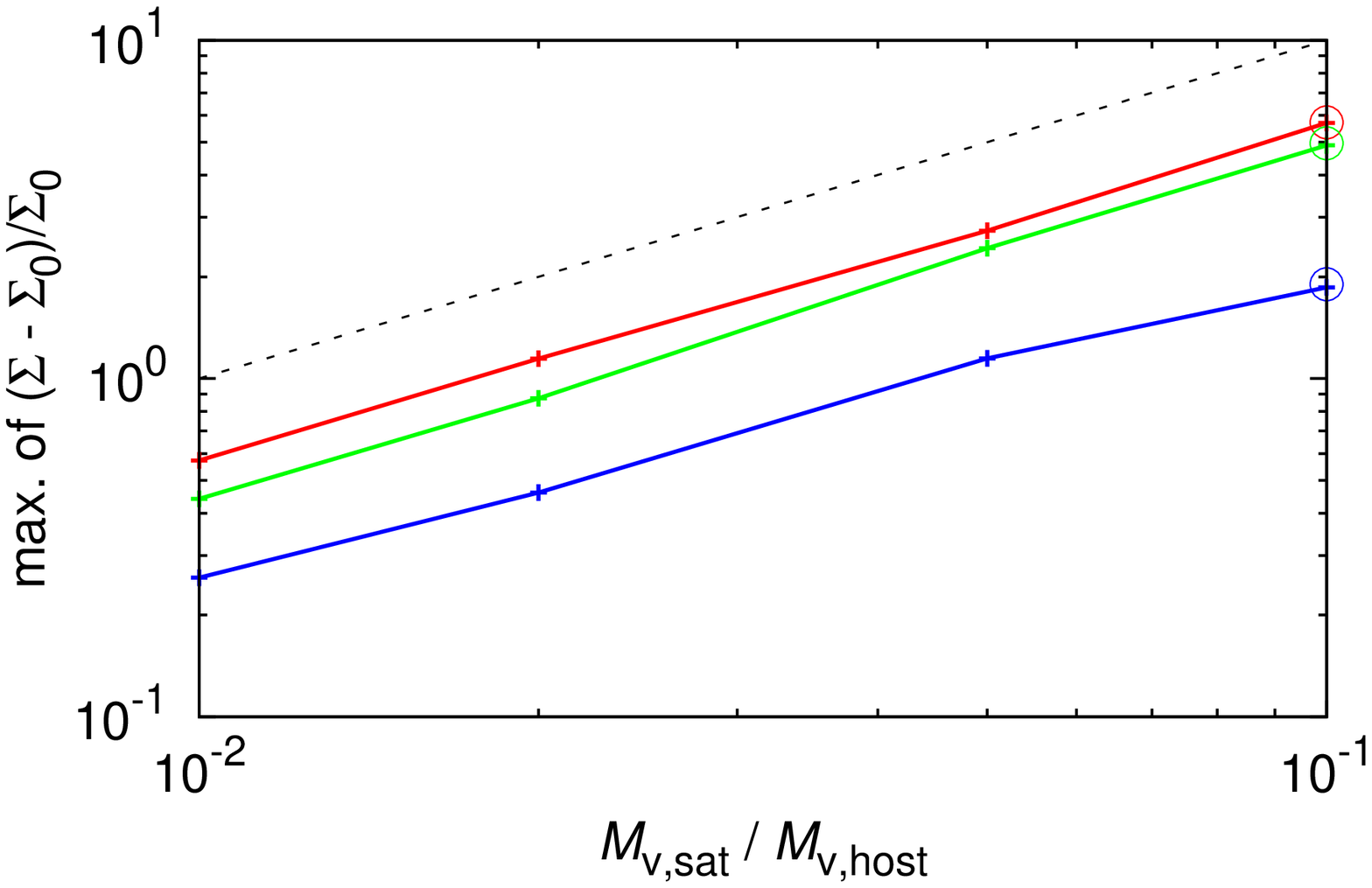}
     \caption{
       Maximum values of enhancements in the column density. 
       Red, green and blue lines show the results for $XY$, $YZ$ and $ZX$ projection planes, respectively. 
       The dashed line represents the scaling, $\propto M_{\rm v,sat}$ which is expected from Equation (\ref{eq:ind_dns}). 
       Crosses correspond to runs A, B, C, D and the open circle is run E. 
       \label{fig:amp}
     }
\end{figure}

We study the relation between the amplitude of the scratches and the satellite mass. 
Figure \ref{fig:amp} represents the maximum value of enhancement in column density as a function of the initial satellite mass, $M_{\rm v, sat}$ and shows that the maximum amplitude is proportional to $M_{\rm v, sat}$. 
The maximum enhancements are obtained when the satellite approaches the first pericenter ($t \sim 1.5 T_{\rm d, host}$) in all simulation runs. 
At that time, the gravitational wake 
merges with the central overdensity 
of the host system. 
In the analytical model, the satellite is regarded as a point mass and the amplitude in the density perturbation is proportional to the satellite mass at given points.  
As a consequence, the amplitude in the column density perturbation should also be proportional to the satellite mass. 
The results of our simulations validate the assumption in the analytical model. 
The comparison of run D and E (see Figure \ref{fig:amp}) shows that the numerical simulations are well converged.

\subsection{Comparison with analytical predictions}
\label{sec:comp}

\begin{figure}
  \centering 
   \includegraphics[width=60mm]{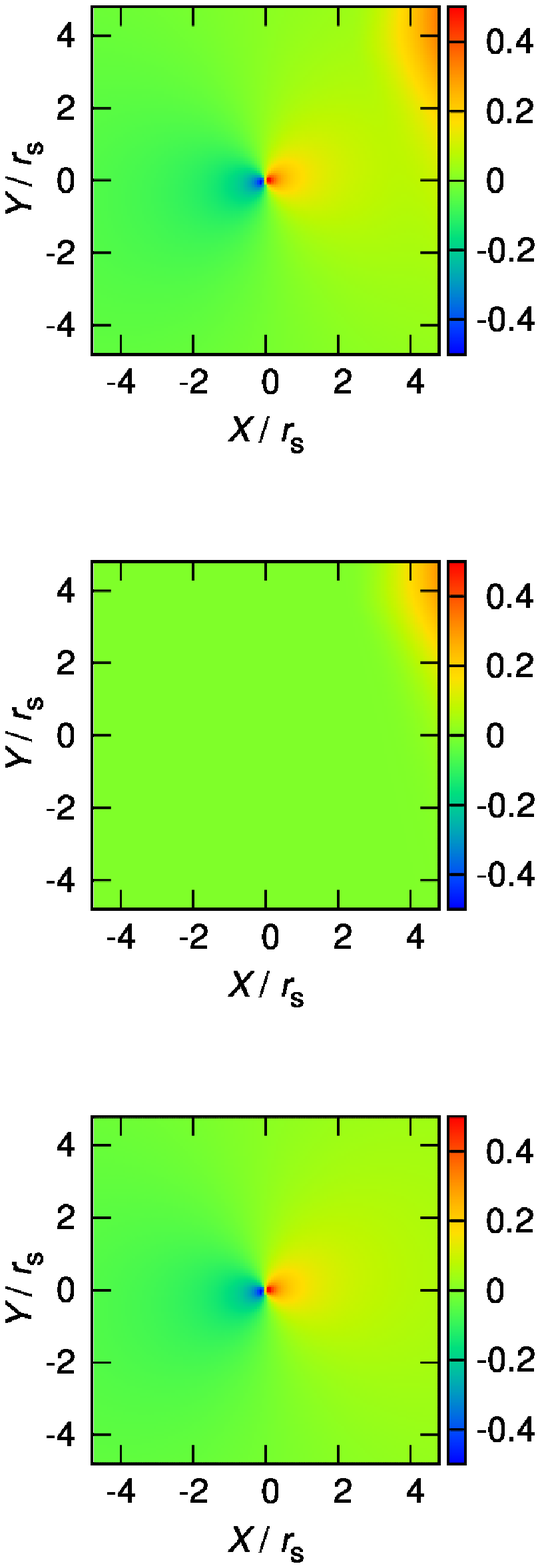}
     \caption{
       Distribution of enhancement and reduction in the column density derived by the analytical model. 
       The position, velocity and mass of the satellite are taken from the snapshot at $t = 1.15 T_{\rm d,host}$ in run E. 
         The top panel 
       demonstrates the total enhancement and reduction. 
       Middle and bottom panels show 
         the 
       contribution from the first and second terms in Equation (\ref{eq:ind_dns}), respectively. 
       Constant numbers, $A = 3.0$ and $B = 1.5$ are adopted and the size of the satellite is set to be $l = r_{\rm s, sat}  = R_{\rm v, sat} / 15$.   
\label{fig:comp}
     }
\end{figure}

In order to test the validity of the analytical model described in Section \ref{sec:ana} and to understand the simulation results, we compare our analytical model predictions with results of the simulation. 
A spherical system which follows an NFW density profile is assumed as the initial unperturbed state. 
This corresponds to assuming that the center of the system is the center of mass. 
Figure \ref{fig:comp} demonstrates the predicted enhancement and reduction in the distribution of column density of the host system. 
Comparing the simulation results, panel (a) in Figure \ref{fig:map1}, with the top panel in Figure \ref{fig:comp}, the analytical prediction well reproduces the results of the simulation, not only the distribution of enhancement and reduction but also the amplitude. 
Middle and bottom panels show the contributions from the first and second terms in Equation (\ref{eq:ind_dns}). 
The results clearly indicate that the first and second terms make the gravitational wake and the mirror image of the over- and underdensities, respectively. 
Because most of previous studies have assumed homogeneous background density, the effects of the second term, the mirror image of the over- and underdensities 
have not been found and discussed. 
However the feature should arise in many astrophysical systems such as galaxies and galaxy clusters since their density distributions have gradients. 

As described above, the predictions well match the simulation results and the analytical model provides a clear understanding. 
However, there still remain small deviations between the simulation results and the analytical predictions. 
The direction of the mirror plane is slightly different. 
This is manly due to two effects. 
First of all, the stripped mass is not considered in the analytical model. 
The stripped mass is distributed along the satellite orbit in the simulation. 
On the other hand, the analytical model treats the satellite as a point mass and does not consider the stripped mass. 
This effect should become more important in the last phase of the merger process since more satellite mass has been stripped. 
The second effect is changes in the density distribution of the host. 
In the analytical model, we assume the initial density distribution of the host as the background field (physical quantities labeled `0' in Section \ref{sec:ana}).  
The density distribution of the host however changes with time. 
Hence, it is sensible to avoid applying the analytical model to systems in violently changing dynamical states. 

\section{Application}
\label{sec:app}
In this Section, we apply the analytical model to nearby interacting galaxy pairs, the Milky Way (MW) - Large Magellanic Cloud (LMC) system and the M31 - M33 system. 
Information about the position, velocity and mass of the satellites, the LMC and M33 and the background density field of the hosts, the MW and M31 are needed. 
We assume an NFW model with a virial mass, $M_{\rm v} = 1.26 \times 10^{12} M_{\rm \odot}$ and concentration parameter $c=10$ for the unperturbed density field of the MW and M31 \citep{2012ApJ...753....8V}. 
Actually, the observed density field of the hosts has been already perturbed, i.e. $\rho_{\rm obs} = \rho_{\rm host} = \rho_0 + \rho_1$. 
We assume here that $|\rho_1|$ is much smaller than $\rho_0$. 
The distance between the solar system and the Galactic Center is assumed to be $R_{\rm sol} = 8.5 {\rm kpc}$ \citep{1986MNRAS.221.1023K}. 
The parameters in Equations (\ref{eq:bmin}) and (\ref{eq:bmax2}) are the same 
as 
those used to plot Figure \ref{fig:comp}, 
$A = 3.0$ and $B = 1.5$.

We take into account only a dark matter halo in the analysis for simplicity. 
Baryon components of host galaxies, such as bulges, discs and stellar haloes also react to the gravitational force of satellite galaxies and density scratches arise in them. 
Ongoing observations, e.g. 
Gaia 
and Subaru Hyper Suprime-Cam., may find not only density fluctuations, but also fluctuations in the velocity field caused by the induced density fields. 
Combining observational data with our 
analytical model might be interesting in order to constrain the the orbits and masses of the satellite galaxies. 

\subsection{MW - LMC}
\label{sec:mw_lmc}
\begin{figure}
  \centering 
   \includegraphics[width=67mm]{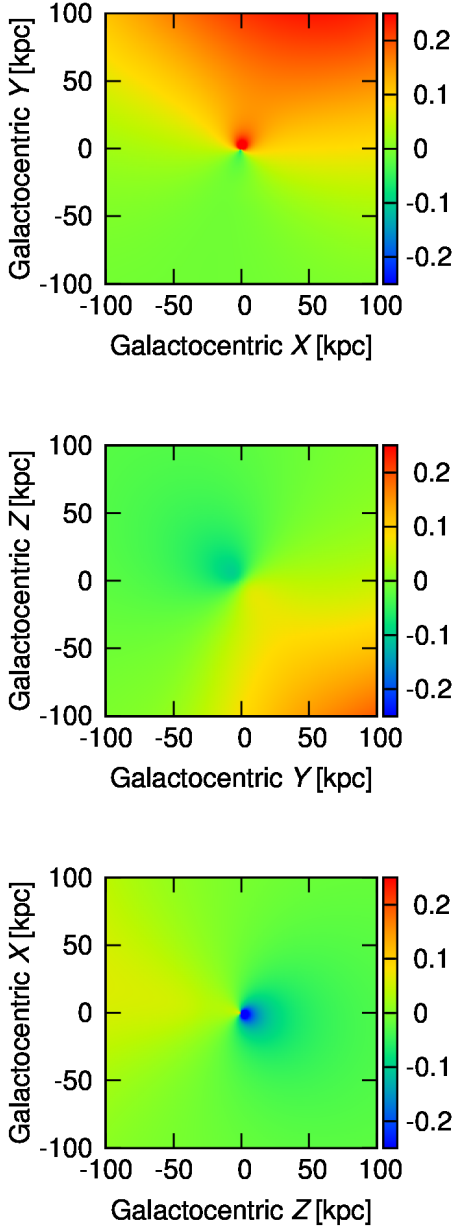}
     \caption{
       Distribution of enhancement and reduction in the column density of the Milky Way induced by the Large Magellanic Cloud. 
       The color bar represents the enhancement and reduction in the column density, $(\Sigma - \Sigma_0) / \Sigma_0$, where $\Sigma_0$ is the column density of the background field at given position in the Galactocentric coordinate. 
       Top, middle and bottom panels show the results for $XY$, $YZ$ and $ZX$ planes, respectively. 
       In each panel, the column density is derived by integration of the density field in the range of $Z = [-200, 0]$ kpc, $X = [-200, -R_{\rm sol}]$ kpc and $Y = [-200, 0]$ kpc. 
       \label{fig:mw_lmc}
     }
\end{figure}

It is useful to adopt a Cartesian coordinate system $(X, Y, Z)$, the so-called Galactocentric rest frame \citep[e.g.][]{1994MNRAS.266..567G}. 
In this coordinate system, the origin corresponds to the Galactic Center and the $X-$, $Y-$ and $Z-$axes point in the direction from the solar system to the Galactic Center, in the direction of the Galactic Rotation of the solar system and 
towards 
the Galactic North Pole, respectively. 
The position of the solar system is given by ${\bf R}_{\rm sol} = (-R_{\rm sol}, 0, 0)$. 
\cite{2002AJ....124.2639V} provide the position of the LMC, 
\begin{equation}
{\bf r}_{\rm LMC} = (-0.78, -41.55, -26.95) \ {\rm kpc}, 
\end{equation}
and the relative volocity of the LMC with respect to the Galactic Center is obtained by \cite{2013ApJ...764..161K}, 
\begin{equation}
{\bf v}_{\rm LMC} = (-57 \pm 13, -226 \pm 15, 221 \pm 19) \ {\rm km \ s^{-1}}. 
\end{equation}

The total dynamical mass of the LMC is uncertain by a factor of 10. 
The enclosed mass within 8.7 kpc from the center of the LMC is $(1.7 \pm 0.7) \times 10^{10} M_{\rm \odot}$ \citep{2014ApJ...781..121V}. 
The total mass should be greater than this value. 
Determining the total mass of the LMC by using the abundance matching technique \citep{2010MNRAS.404.1111G}, the upper mass limit of the LMC is $2.5 \times 10^{11} M_{\rm \odot}$ \citep{2013ApJ...764..161K}. 
This is consistent with the estimation by \cite{2015arXiv150703594P}. 
We assume that the mass and size of the LMC are $M_{\rm LMC} = 10^{11} M_{\rm \odot}$ and $l = 8.7 {\rm kpc}$, respectively. 
As shown in Figure \ref{fig:amp}, the amplitude of 
the 
density enhancement can be scaled by $\propto M_{\rm LMC}$. 

Figure \ref{fig:mw_lmc} demonstrates the predicted enhancement and reduction in the column density distribution of the MW. 
$XY$ and $ZX$ planes in the Galactocentric coordinate are good to find clear density scratches of the LMC. 
When one sees the south-side sky, the column density in the direction of the Galactic Rotation of the solar system (plus $Y$) is expected to be systematically greater than that in the opposite direction (upper panel). 
Also, the column density on the side of the Galactic North Pole (plus $Z$) should be systematically lower than that on the opposite side when one looks into the opposite direction of the the Galactic Rotation of the solar system (lower panel). 

\subsection{M31 - M33}
\label{sec:m31_m33}
\begin{figure}
  \centering 
  \includegraphics[width=57mm]{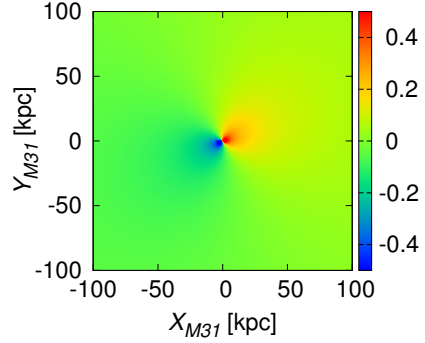}
     \caption{
       Distribution of enhancement and reduction in the column density of M31 induced by M33. 
       The color bar represents the enhancement and reduction in the column density, $(\Sigma - \Sigma_0) / \Sigma_0$, where $\Sigma_0$ is the column density of the background field at given position in the M31 rest frame. 
       The column density is derived by integration in the range of $Z_{\rm M31} = [-200, 200]$ kpc. 
       \label{fig:m31_m33}
     }
\end{figure}

The position and velocity vectors of M31 and M33 are obtained by \cite{2012ApJ...753....8V}. 
We adopt a coordinate system, $(X_{\rm M31}, Y_{\rm M31}, Z_{\rm M31})$ in which the origin is the center of M31 and the $Z_{\rm M31}-$axis points in the direction from the solar system to the center of M31.  
$X_{\rm M31}-$ and $Y_{\rm M31}-$axes are perpendicular to the $Z_{\rm M31}-$axis and a right‐handed system is constructed. 
In the M31 rest frame, the position and velocity of M33 are  
\begin{eqnarray}
{\bf r}_{\rm M33} &=& (140.5, 146.1, -2.3) \ {\rm kpc}, \nonumber \\ 
{\bf v}_{\rm M33} &=& (-147.4, -72.2, 117.9) \ {\rm km \ s^{-1}}.  
\end{eqnarray}

The total mass of M33 is uncertain by a factor of 10 
similar to the LMC mass. 
\cite{2003MNRAS.342..199C} found that dark halo mass out to 17 kpc from the center of M33 is $\sim 5 \times 10^{10} M_{\rm \odot}$. 
\cite{2011ISRAA2011E...4S} obtained 
a 
virial mass of the dark halo surrounding M33 
of 
$(2.2 \pm 0.1) \times 10^{11} M_{\rm \odot}$ from the HI rotation curve. 
We assume that the mass and size of M33 are $M_{\rm M33} = 10^{11} M_{\rm \odot}$ and $l = 17 {\rm kpc}$, respectively.  

The analytical model 
predicts a mirror image of the density enhancement and reduction around the center of M31 as shown in Figure \ref{fig:m31_m33}. 
The result 
appears to violate 
the basic assumption, $\rho_0 \gg |\rho_1|$, but the amplitude of density enhancement can be scaled by $\propto M_{\rm M33}$. 

\section{Summary and discussion}
\label{sec:sum}

We have investigated the dynamical response of extended host systems to the gravitational force of orbiting satellite systems, `scratches'. 
The scratches are classified into two types: 
the first one is the gravitational wake along the orbit of satellites as discussed in \cite{1943ApJ....97..255C}. 
The second type is a mirror image of the over- and underdensities which become more evident in the center of the hosts. 
The mirror plane is perpendicular to the direction of the satellite velocity. 
We derive features analytically from Chandrasekhar's formula of dynamical friction. 
Our $N$-body simulations validate the analytical predictions well. 

The scratches may be found in nearby interacting galaxies by observations. 
The dynamical mass including a dark halo of the satellite galaxies, the LMC and M33 is still uncertain by a factor of 10 \citep[e.g.][]{2003MNRAS.342..199C, 2011ISRAA2011E...4S, 2013ApJ...764..161K, 2014ApJ...781..121V}. 
As indicated by Equation (\ref{eq:ind_dns}) and shown in Figure \ref{fig:amp}, the amplitude of the induced density is 
proportional 
to the satellite mass. 
Combining the analytical model with observations, new constraints for the satellite masses may be provided. 

The form of the Coulomb logarithm is important in order to determine the features and amplitudes of scratches. 
In this paper, we adopt a simple formula to provide the Coulomb logarithm as a function of position. 
A caveat is the constant minimum impact parameter, $b_{\rm min}$ in Equation (\ref{eq:bmin}) and the parameter, $B$ in Equation (\ref{eq:bmax2}). 
We determine them by fitting analytical predictions to the simulation result but they may vary from system to system. 
Actually, $b_{\rm min}$ may depend on the local density since it should have similar values as the typical distance between the satellite and nearby particles. 
More systematic studies can help to improve the form of the Coulomb logarithm. 

In a later step it also would be important to consider more realistic 
configurations 
of host systems with many satellite systems orbiting around them. 
The analytical arguments in this paper might also help to understand the dynamical phenomena in 
these 
more complexed systems. 

\section*{Acknowledgments}
We are grateful to the anonymous referee for providing many helpful comments and suggestions. 
We thank Alessandro Ballone, Manuel Behrendt, Masashi Chiba, Jorge Cuadra, Guinevere Kauffmann, Takanobu Kirihara, Lucio Mayer, Yohei Miki, Masao Mori, Daisuke Nagai, Simon White and Kohji Yoshikawa for fruitful discussions. 
Numerical simulations were performed with HA-PACS at the Center for Computational Sciences at University of Tsukuba. 
This work was supported by Grant-in-Aid for JSPS Fellows (25-1455 GO) and the DFG cluster of excellence `Origin and Structure of the Universe' (www.universe-cluster.de).

\bibliographystyle{mn2e}
\bibliography{./ref}

\label{lastpage}
\end{document}